\documentclass[twoside]{dis08}
\usepackage[latin1]{inputenc}
\usepackage[dvips]{graphicx,epsfig,color}
\usepackage{wrapfig,rotating}
\usepackage{amssymb,amsmath,array}

\def\wzjets{$W^{\pm}/Z+{\rm jets}$}
\def\wjets{$W+{\rm jets}$}
\def\zjets{$Z/\gamma^{*}+{\rm jets}$}
\def\zele{$Z/\gamma^{*}\rightarrow e^{+}e^{-}$}
\def\wele{$W^{\pm}\rightarrow e^{\pm}\nu$}
\def\wtau{$W\rightarrow \tau\nu$}
\def\welejets{$W\rightarrow e^{\pm}\nu$ + jets}

\def\pt{$P_{T}$}
\def\et{$E_{T}$}
\def\ppbar{$p\overline{p}$}

\def\ttbar{$t\overline{t}$}
\def\wnjets{$W\rightarrow e\nu + \geq n {\rm ~jets}$}
\def\znjets{$Z/\gamma^{*}\rightarrow e^{+}e^{-} + \geq n {\rm ~jets}$}

\begin{document}
\title{Measurements of Vector Bosons Produced in Association with Jets}

\author{Ben Cooper \\
\em{on behalf of the CDF and D0 Collaborations}
%
%
\vspace{.3cm}\\
%
Queen Mary University of London - Department of Physics \\
London - UK\\
%
}

\maketitle

\begin{abstract}
The latest D0 and CDF measurements of the \wjets~and \zjets~processes are described, along with a discussion of the 
comparisons that have been made to LO and NLO perturbative QCD predictions. 
\end{abstract}

\section{Introduction}

The direct production of $W^{\pm}/Z$ bosons in association with jets is a process of 
crucial importance at hadron collider experiments. The presence of a vector boson in the 
hard scatter means that these interactions occur at a scale that should make perturbative 
QCD applicable, and thus it is an excellent channel to test such predictions. Furthermore, many of the 
potential discovery channels for the Higgs boson and beyond standard model processes share a final state signature 
with the \wzjets~process. It is thus vital for the success of existing and 
future hadron collider experiments that this process is understood, and recently there has been a huge 
amount of work put into the modelling of this process, with the appearance of 
many new Monte Carlo generators that are already widely used at both the Tevatron and LHC. In Sections~\ref{Sec:zjets} 
and~\ref{Sec:wjets} the latest \wjets~and \zjets~measurements from the Tevatron are presented, and in Section~\ref{Sec:th} 
we discuss the results and implications of some of the theory comparisons that have thus far been made.

\section{\zjets~Measurements}
\label{Sec:zjets}


Both the CDF and D0 collaborations have produced \zjets~measurements in the \zele~channel~\cite{CDFZ, D0Z}, using $\rm{1.7}fb^{-1}$ 
and $\rm{400}pb^{-1}$ of Tevatron Run II data respectively. D0 has measured the ratio of \znjets~production cross sections to the 
total inclusive \zele~cross section for $n = 1 - 4$ and jet $P_{T} > {\rm 20~GeV}$. CDF has measured the inclusive \znjets~differential 
cross section as a function of jet \pt~for $n = 1,2$ and $P_{T} > {\rm 30~GeV}$. In both measurements, \zele~events are selected by requiring 
two electrons with $P_{T} > {\rm 25~GeV}$ that together form an invariant mass compatible with a 
Z resonance. In the D0 analysis only electrons within the central region 
of the calorimeter were used, whereas CDF used one central electron and allowed the second one to be either in the central or 
forward region of the calorimeter.

Both analyses use ``tag and probe'' methods to extract from the data efficiencies for electron identification, and both correct 
for the acceptance of the kinematic and geometrical selection criteria by using simulated signal Monte Carlo samples. In the CDF 
analysis, the measured cross section is defined for a limited kinematic range of the \zele~decay products (corresponding to the event 
selection criteria), and the acceptance factor is defined accordingly. In this way, the sensitivity of the measurement to the 
theoretical modelling of the signal is reduced. 
\begin{wrapfigure}[18]{L}{0.50\textwidth}
\centerline{\includegraphics[width=0.50\textwidth]{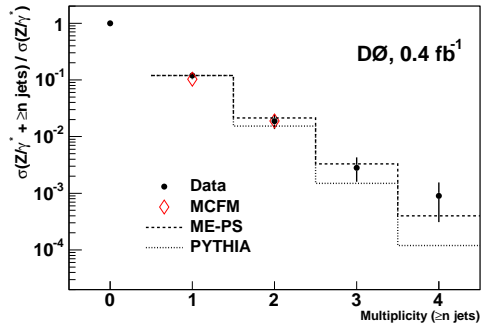}}
\caption{The D0 measured \znjets/\zele~cross section ratios for $n = 1-4$ compared to three different predictions: {\sc mcfm} NLO, 
a ME-PS matched prediction and {\sc pythia}.}
\label{Fig:D0Z}
\end{wrapfigure}

The dominant sources of background to the \zjets~process are those arising from QCD multijet events and \wjets~events. In the CDF analysis a 
data-driven method is used to estimate both these sources by extracting from data the probability for an additional
jet to fake the second leg of a Z boson decay in events with one electron in the final state. In the D0 analysis the QCD background 
is extracted from an analysis of the sidebands of the Z peak, and the \wjets~background is taken from simulated Monte Carlo. In the 
D0 case the total background is found to be 3-5\%, whereas in the CDF analysis it is at the level of 12-17\%.

%

In the D0 measurement, 
jets were clustered using a cone algorithm~\cite{D0ZJET} with cone radius $R = 0.5$, requiring jet $P_{T} > {\rm 20~GeV}$ and $|\eta| < 2.5$. 
In the CDF measurement, jets 
were clustered using a seeded midpoint cone algorithm with cone radius $R = 0.7$, requiring jet
$P_{T} > \rm{30~GeV}$ and $|y|<2.1$. In both analyses data-driven methods were used to correct the transverse momentum of the jets to account for multiple 
\ppbar~interactions and the response of the calorimeter. 
In addition, jet reconstruction and identification efficiencies as well as the impact of the finite jet energy resolution 
of the detector are accounted for in the cross sections using simulated Monte Carlo samples that have been tuned on data. 

In both CDF and D0 analyses the dominant source of systematic uncertainty is that arising from the determination of the 
calorimeter jet energy scale. In the CDF measurement the total systematic is $\sim10\%$ (15\%) at low (high) jet~\pt. In the D0 
measurement the total systematic is similar for $n \geq 1,2$ but reaches $\sim50\%$ for $n \geq 4$.

\section{\wjets~Measurements}
\label{Sec:wjets}

The CDF collaboration has recently published a \wjets~measurement~\cite{CDFW} in the \wele~channel using $\rm{320}pb^{-1}$ of 
Tevatron Run II data, measuring the differential \wnjets~cross section as a function 
of the $n$th highest \et~jet above 20 GeV, for $n = 1 - 4$. 
\begin{wrapfigure}[26]{R}{0.45\textwidth}
\centerline{\includegraphics[width=0.45\textwidth]{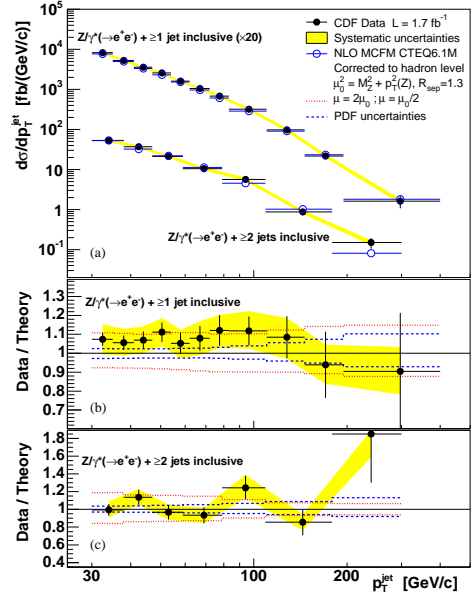}}
\caption{The CDF measured inclusive \znjets~differential cross sections as a function of jet \pt~ 
for events with $n \geq 1,2$, compared to NLO {\sc mcfm} predictions that have been corrected to the 
hadron level.}\label{Fig:CDFZ}
\end{wrapfigure}
In this analysis \welejets~events were selected by 
requiring exactly one central electron  with $P_{T} > {\rm 20~GeV}$ along with missing transverse energy $E\!\!\!/_{T} > {\rm 30~GeV}$. 
It was also required that the reconstructed $W$ transverse mass satisfies $M_{T}^{W} > 20~GeV/c^{2}$, a cut which reduces background 
at little expense to the signal. 
As in the CDF \zjets~analysis, the cross section is defined for a limited kinematic range of the $W$ decay 
products equal to this event selection criteria, and the acceptance and efficiencies are computed from simulated signal Monte 
Carlo samples accordingly.

Jets in the \wele~event sample were clustered using the {\sc jetclu}~\cite{JETCLU} cone algorithm with cone size 0.4, requiring jet $E_{T} > {\rm 20~GeV}$ and 
$|\eta| < 2.0$. The energy of each jet is corrected for multiple interactions and the calorimeter response. 
In addition, once the jet spectra had been corrected for backgrounds, simulated Monte Carlo signal samples were used to correct the jet spectra 
to account for the jet reconstruction efficiency and finite calorimeter energy resolution.

The dominant sources of background to the \wjets~process arise from from QCD multijet and \ttbar~production. In this analysis, the multijet background was 
estimated by using an alternative ``antielectron'' selection criteria to select from the data an event sample that could be used to reliably 
model the QCD background in the required jet kinematic distributions. The background from \ttbar, as well as the less important \wtau, \zele, $WW$ and $W\gamma$
processes, was modelled using simulated Monte Carlo samples. 
The total background fraction increases with increasing jet multiplicity and transverse energy, around 10\% at low \et, but reaching 90\% at the highest 
measured \et~in the four jet sample. 

At low jet \et~the dominant systematic on the measured 
cross sections is that arising from the jet energy scale determination, at the level of 5-10\%. However, at higher jet \et~the uncertainty on the background 
determination is dominant, up to 80\% in the highest jet \et~bins.

\section{Theoretical Comparisons}
\label{Sec:th}

\begin{wrapfigure}[25]{R}{0.45\textwidth}
\centerline{\includegraphics[width=0.45\textwidth]{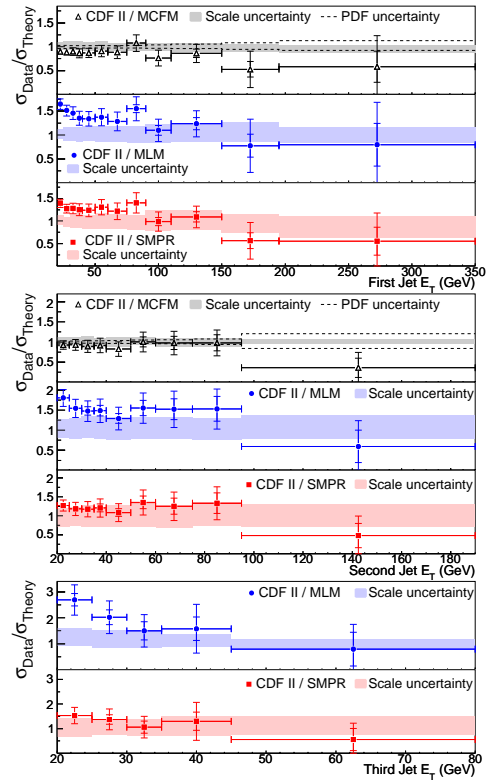}}
\caption{The CDF measured \wnjets~differential cross section for $n = 1-3$ compared to {\sc mcfm} NLO and ME-PS matched predictions.
}\label{Fig:CDFW}
\end{wrapfigure}
Figures~\ref{Fig:D0Z},\ref{Fig:CDFZ} and~\ref{Fig:CDFW} show the results of the measurements described above compared to various next-to-leading order 
(NLO) and leading order (LO) perturbative QCD predictions. All three measurements make comparisons to {\sc mcfm}~\cite{MCFM} NLO predictions with up to 2 partons in the 
final state. These calculations are made at the parton level, and as such do not include the effects of hadronization or the underlying event which will 
be present in the data. 
However, the CDF Z measurement uses a {\sc pythia tune-a}~\cite{PYTHIA, TUNEA} Monte Carlo 
sample to derive for each jet \pt~bin a parton-to-hadron correction factor that is applied to the {\sc mcfm} predictions to approximately account for these 
non-perturbative contributions. In Figure~\ref{Fig:CDFZ} the CDF inclusive \znjets~differential cross sections as a function of jet \pt~are compared 
with the corrected {\sc mcfm} predictions, and good agreement is observed between the data and the prediction both in terms of overall rates and in the 
reproduction of the spectra shape. However, in Figures~\ref{Fig:D0Z} and~\ref{Fig:CDFW} one can see that the {\sc mcfm} prediction still well reproduces 
the data even in the absence of such corrections. In the CDF \wjets~analysis it was observed that, for this particular jet definition, 
the effects of hadronization and the underlying event cancel each other out at the 5-10\% level~\cite{CDFW}.

In Figure~\ref{Fig:D0Z} comparisons of the D0 \zjets~data are made to a LO matrix element parton shower matched prediction~\cite{MEPS} (ME-PS) 
based on a modified CKKW scheme~\cite{CKKW,MRENNA}, and to {\sc pythia}. These predictions have been normalised to the measured $Z/\gamma^{*}+\geq 1$ 
jet cross section ratio. One can see that the ME-PS matched predictions better reproduce the rate of additional jets due to the inclusion of tree-level 
processes of up to three partons. 

In Figure~\ref{Fig:CDFW} comparisons of the CDF \wjets~data are made to two different  ME-PS matched predictions; 
{\sc madgraph}~\cite{MADGRAPH} + {\sc pythia} predictions using a modified CKKW scheme~\cite{MRENNA} (SMPR), and {\sc alpgen}\cite{ALPGEN} + 
{\sc herwig}~\cite{HERWIG} predictions using the MLM scheme~\cite{MLM} (MLM). These predictions are not normalised to the data, and the limitations of LO calculations 
in reproducing absolute rates is clear. However, the SMPR prediction in particular well reproduces the shape of the measured spectra. 
The discrepancies 
observed at low jet \et~in the comparison to MLM are possibly due to the absence of a Tevatron Run II tuned underlying event model in this prediction.


\section{Summary}
\label{Sec:sum}

The recent CDF and D0 measurements of the \wzjets~process open the door for a thorough exploration into the 
ability of the latest theoretical predictions to model this important process. 
The comparisons to theory that have been made thus far indicate that important and impressive progress has been made with the latest 
Monte Carlo generators.

\vspace{0.9cm}


\begin{footnotesize}

\end{footnotesize}


\end{document}